\documentclass[]{emulateapj}

\newcommand{\lsim}{\lower0.6ex\vbox{\hbox{$ \buildrel{\textstyle <}\over{\sim}\ $}}}
\newcommand{\gsim}{\lower0.6ex\vbox{\hbox{$ \buildrel{\textstyle >}\over{\sim}\ $}}}

\newcommand{\hmpc}{h^{-1}\mathrm{Mpc}}
\newcommand{\hkpc}{h^{-1}\mathrm{kpc}}

\newcommand{\hMsun}{\ h^{-1}\mathrm{M}_{\odot}}
\newcommand{\hMpc}{\ h^{-1}\mathrm{Mpc}}
\newcommand{\Msun}{M_{\odot}}
\newcommand{\kms}{{\,{\mathrm{km}}\,{\mathrm{s}}^{-1}}}

\newcommand{\rhomean}{\rho_{\mathrm{M}}}

\newcommand{\mpt}{m_{\mathrm{p}}}
\newcommand{\rfind}{r_{\mathrm{f}}}

\newcommand{\Mvir}{M_{\mathrm{vir}}}
\newcommand{\Rvir}{R_{\mathrm{vir}}}
\newcommand{\Dvir}{\Delta_{\mathrm{vir}}}
\newcommand{\Vmax}{V_{\mathrm{max}}}
\newcommand{\Vhost}{V_{\mathrm{host}}}
\newcommand{\Vsat}{V_{\mathrm{sat}}}

\newcommand{\Nsat}{N_{\mathrm{sat}}}
\newcommand{\Ngal}{N_{\mathrm{gal}}}

\def\kms{{\ }{\rm km}\,{\rm s}^{-1}}
\def\LCDM{$\Lambda$CDM}

\newcommand{\cvir}{c_{\mathrm{vir}}}
\newcommand{\ac}{a_{\mathrm{c}}}

\newcommand{\mymcf}[1]{\mathcal{M}_{#1}}
\newcommand{\mymcfc}{\mathcal{M}_{c_{\mathrm{vir}}}}
\newcommand{\mymcfa}{\mathcal{M}_{a_{\mathrm{c}}}}
\newcommand{\mymcfN}{\mathcal{M}_{N_{\mathrm{sat}}}}

\newcommand{\cmark}{\ifmmode{\widetilde{c}_{\mathrm{vir}}}\;\else $\widetilde{c}_{\mathrm{vir}}$\fi}
\newcommand{\amark}{\widetilde{a}_{\mathrm{c}}}
\newcommand{\nmark}{{N_{\mathrm{sat}}}}

\newcommand{\var}[1]{\mathcal{V}(#1)}

\newcommand{\avg}[1]{\langle #1 \rangle}
\newcommand{\pairavg}[1]{\langle #1 \rangle_{\mathrm{p}}}

\newcommand{\zobs}{\ifmmode{z_{\rm o}}\else$z_{\rm o}$\fi}
\newcommand{\aobs}{\ifmmode{a_{\rm o}}\else$a_{\rm o}$\fi}
\newcommand{\Mobs}{\ifmmode{M_{\rm o}}\else$M_{\rm o}$\fi}       

\newcommand{\omegam}{\Omega_{\mathrm{M}}}
\newcommand{\omegal}{\Omega_{\Lambda}}
\newcommand{\sig}{\sigma_{8}}
\newcommand{\mstar}{M_{\star}}
\newcommand{\brel}{b_{c_{\mathrm{vir}}}}

\newcommand{\dd}{\mathrm{d}}

\newcommand{\beq}{\begin{equation}}
\newcommand{\eeq}{\end{equation}}
\newcommand{\beqa}{\begin{eqnarray}}
\newcommand{\eeqa}{\end{eqnarray}}

\newcommand{\cp}{c^\prime}
\newcommand{\mmark}{\widetilde{m}}

\newcommand{\xith}{\xi^{2\mathrm{h}}}
\newcommand{\xioh}{\xi^{1\mathrm{h}}} 
\newcommand{\xihh}{\xi_{\mathrm{hh}}}  
\newcommand{\ngal}{\overline{n}_{\mathrm{g}}}   
\newcommand{\bh}{b_{\mathrm{h}}}    
\newcommand{\xidm}{\xi_{\mathrm{dm}}}  
                          
\shortauthors{WECHSLER et al}
\shorttitle{THE DEPENDENCE OF HALO CLUSTERING ON CONCENTRATION}
\begin{document}
\journalinfo{The Astrophysical Journal, in press}
\submitted{Received 2005 December 15; accepted 2006 June 23}

\title{
The Dependence of Halo Clustering on Halo Formation History, \\Concentration, and Occupation
}

\author{
Risa H. Wechsler\altaffilmark{1,2}, 
Andrew R. Zentner\altaffilmark{1},
James S. Bullock\altaffilmark{3},
Andrey V. Kravtsov\altaffilmark{1}
Brandon Allgood\altaffilmark{4}
}
\altaffiltext{1}{
  Kavli Institute for Cosmological Physics,
  Department of Astronomy \& Astrophysics, and 
  Enrico Fermi Institute,
  The University of Chicago, 
  Chicago, IL 60637 USA
}
\altaffiltext{2}{
  Hubble Fellow
}
\altaffiltext{3}{ 
  Center for Cosmology,
  Department of Physics \& Astronomy, 
  University of California,
  Irvine, CA 92697 USA
}
\altaffiltext{4}{
  Physics Department,
  University of California,
  Santa Cruz, CA 96050 USA; present address: 
  Pharmix Corp., 2000 Sierra Point Pkwy, Suite 500, Brisbane, CA 94005
}
\begin{abstract}
We investigate the dependence of dark matter halo clustering on halo
formation time, density profile concentration, and subhalo occupation
number, using high-resolution numerical simulations of a LCDM
cosmology.  We confirm results that halo clustering is a function of
halo formation time at fixed mass, and that this trend depends on halo
mass.  For the first time, we show unequivocally that halo clustering
is a function of halo concentration and show that the dependence of
halo bias on concentration, mass, and redshift can be accurately
parameterized in a simple way: $b(M,c|z) = b(M|z) \brel(c|M/M_*)$.
Interestingly, the scaling between bias and concentration changes sign
with the value of $M/M_*$: high concentration (early forming) objects
cluster more strongly for $M \lsim M_*$, while low concentration (late
forming) objects cluster more strongly for rare high-mass halos, $M
\gsim M_*$.  We show the first explicit demonstration that host dark
halo clustering depends on the halo occupation number (of dark matter
subhalos) at fixed mass, and discuss implications for halo model
calculations of dark matter power spectra and galaxy clustering
statistics.  The effect of these halo properties on clustering is
strongest for early-forming dwarf-mass halos, which are significantly
more clustered than typical halos of their mass.  Our results suggest
that isolated low-mass galaxies (e.g. low surface-brightness dwarfs)
should have more slowly-rising rotation curves than their more
clustered counterparts, and may have consequences for the dearth of
dwarf galaxies in voids.  They also imply that self calibrating
richness-selected cluster samples with their clustering properties
might overestimate cluster masses and bias cosmological parameter
estimation.
\end{abstract}


\section{Introduction}
\label{section:intro}

The spatial distribution of galaxies is now well established to be
dependent on several of their internal properties: stellar mass,
luminosity, color, star formation rate, Hubble type, and several
others
\citep[e.g.,][]{hubble:36,dressler:80,norberg_etal:01,zehavi_etal:02}.
In the current paradigm for galaxy formation, this can be understood
as a combination of the fact that dark matter halos with different
masses and formation histories cluster differently and host different
galaxy populations.  A full understanding of these trends is one of
the primary goals of modern cosmology, as it is likely to provide
insight into the physical process that govern galaxy formation and aid
in the use of observed galaxy clustering as a probe of fundamental
cosmological parameters.

Many models of galaxy formation and methods of calculating galaxy
clustering statistics make two related simplifying assumptions.  The
first is that the number and properties of galaxies within a host dark
matter halo depend solely on the mass of the host, independent of halo
environment or other properties of the dark halo.  The second is that
the clustering properties of dark matter host halos are a function
only of their masses and that dark matter halos are otherwise ignorant
of their larger environments.  The latter assumption forms part of the
basis of the excursion-set formalism for galaxy clustering
\citep{bond_etal:91}, at least in its simplest and most common
implementation \citep{lacey_cole:93,somerville_kolatt:99}.  This
implementation assumes that halo formation is a Markov process with no
correlations between different spatial scales, which then implies that
future halo accretion is independent from past history, and that halo
histories are independent of environment (see also discussion in
\citealt{white:96} and \citealt{sheth_tormen:04}).

The first place that these assumptions are made is in semi-analytic
models of galaxy formation.  A basic assumption of the technique is
that the properties of galaxies depend only on the mass and formation
time of the host halo.  In many implementations, galaxy clustering is
calculated by filling simulated halos at a fixed redshift with halo
formation histories that are calculated analytically using the
excursion-set formalism
\citep[e.g.][]{kauffmann_etal:97,benson_etal:00,wechsler_etal:01,zentner_etal:05}.
By construction, any dependence of galaxy properties on environment in
these models must come only from the extent to which they populate
halos of different masses.  Note that this assumption is avoided to
some extent in many modern implementations which use halo merging
histories extracted directly from N-body simulations
\citep[e.g.][]{springel_etal:01, helly_etal:03, hatton_etal:03,
springel_etal:05, kang_etal:05, croton_etal:05, delucia_etal:06,
bower_etal:05}.  In these implementations semi-analytic recipes depend
on mass and formation history as before, but correlations between the
galaxy properties and environment may now come from the extent to
which they populate halos of different masses {\em and formation
histories}.  If there are physical effects that otherwise depend on
the larger scale environment these would not be included.

The second place these assumptions are commonly made is in the
standard halo model of galaxy clustering (e.g.,
\citealt{seljak:00,peacock_smith:00,scoccimarro_etal:01};
\citealt*{berlind_weinberg:01,bullock_etal:02};
\citealt{cooray_sheth:02}).  The halo model is a framework for
calculating clustering statistics of objects by associating them with
dark matter halos, which have well-studied abundances and clustering
properties.  The clustering statistics of a galaxy population can be
computed after specifying the clustering properties of the dark matter
halos, the probability distribution for the number of galaxies in a
host halo as a function of halo mass, and the distribution of these
galaxies within their host halos.  In principle, the halo bias, the
probability distribution for the number of galaxies in a halo of fixed
mass, and the spatial distribution of galaxies within their host halos
can depend on other properties of the halo, but the standard
assumption is that they depend only on halo mass.
\citet{abbas_sheth:05} have recently described a modification of the
halo model that incorporates dependencies on local densities.

\citet{lemson_kauffmann:99} tested the assumption that halo properties
are independent of environment using numerical simulations of
cosmological structure formation and found no dependence of halo
clustering on formation time or on several other properties of the
dark halos.  More recent theoretical studies have also indicated that
any trends of halo occupation on environment have only a relatively
small net effect on large-scale clustering statistics, at least at the
level that can be measured in relatively small computational volumes
\citep{berlind_etal:03,zentner_etal:05,yoo_etal:05}.  However, a
recent study by \citet{avila-reese_etal:05} found environmental trends
with halo concentration, spin, shape, and internal angular momentum.

An early indication of a relationship between halo formation histories
and halo clustering properties was demonstrated by \citet[][see also
\citealt{wechsler:01}]{sheth_tormen:04}.  Recently,
\citet*{gao_etal:05} showed convincingly that the clustering of
low-mass halos is a strong function of their formation times.  In this
mass regime, early-forming halos are significantly more clustered than
their late-forming counterparts.  \citet{harker_etal:06} provided
confirmation of these results using statistics of marked point
distributions similar to those that we employ below.  The trend they
identified is strongest for low-mass halos, which have only recently
been well resolved in numerical simulations.

Many properties of dark matter halos correlate well with halo
formation time, so it is natural to investigate whether these trends
with formation time extend to other halo properties.  In particular,
it is natural to expect trends with the concentrations of dark halo
density profiles, which \citet*{navarro_etal:97},
\citet{wechsler_etal:02}, and \citet{zhao_etal:03} have shown to
correlate well with halo formation time.  In addition, several studies
have convincingly demonstrated that the number of satellite halos
within a host halo of fixed mass is a function of halo formation time
\citep[][]{gao_etal:04,zentner_etal:05,vandenbosch_etal:05,taylor_babul:05}
and halo concentration \citep[][]{zentner_etal:05}.  If satellite
halos are to be associated with satellite galaxies in groups and
clusters, this indicates that the probability distribution for the
number of galaxies per halo, known as the halo occupation distribution
(HOD), is also a function of these variables and, by extension, may
likely be a function of halo environment.

Given the correlations between these halo properties and formation
time, it is interesting to determine whether or not they relate to
environment in a similar way.  This does not have to be the case; if
the relations between these halo properties and formation time are
themselves a function of environment, their trends with clustering
could in principle be quite different.  Moreover, as we discuss below,
concentration and halo occupation have much more direct consequences
for the halo model and its application to constraints on cosmological
parameters, and may have a more direct impact on tests of galaxy
clustering.  We focus our study on these halo properties.

In the present study we use two large, high-resolution dissipationless
cosmological simulations to study the dependence of the clustering of
dark matter halos on halo properties other than mass.  We show that
halo clustering depends on halo formation time, and present a clear
demonstration that halo clustering is a function of both halo
concentration and halo occupation number.  We show how these
properties change with halo mass, and present the first investigation
into how they change with redshift.  We present a simple fitting
formula for our concentration-dependent clustering results that will
enable estimates of the strength of these effects for various
applications in the context of the halo model.  We discuss several
implications of these results for outstanding issues in galaxy
formation, including the clustering of dwarf galaxies and the
concentrations of low surface-brightness galaxies, and for the
estimation of cosmological parameters, including self-calibration of
cluster masses.

We begin with a description of our methods in
\S~\ref{section:methods}.  Specifically, we describe our numerical
simulations in \S~\ref{sub:sim}, and the statistics that we employ in
\S~\ref{sub:mcf}.  In \S~\ref{sub:cvir} and \S~\ref{sub:ac} we discuss
our definitions and measurements of halo concentration and halo
formation time respectively.  In \S~\ref{section:results}, we explore
the clustering dependence of both halo formation time and halo
concentration, and present a model for the relative bias of halos as a
function of concentration.  In this section we also update previous
results on the correlations between both halo formation time and halo
concentration and the number of satellite halos contained within a
host halo of fixed mass.  Following this, we give the first explicit
demonstration that host halo clustering is a function of the
occupation number of satellite halos.  In
\S~\ref{section:implications}, we discuss the implications of our
results for the halo model.  We conclude with a summary of our primary
results and a discussion of their implications for galaxy formation
models and for cosmological constraints derived from galaxy clustering
in \S~\ref{section:disc}.

\section{Methods}
\label{section:methods}

\subsection{Numerical Simulations}
\label{sub:sim}

We investigate the environmental dependence of halo concentrations and
halo occupation using cosmological $N$-body simulations of structure
formation in the concordance, flat $\Lambda$CDM cosmology with
$\omegam = 1 - \omegal = 0.3$, $h=0.7$, and $\sig = 0.9$.  The
simulations were performed with the Adaptive Refinement Tree (ART)
$N$-body code \citep{kravtsov_etal:97}.  The two simulations follow
the evolution of $512^3$ particles in computational boxes of size $120
\hmpc$ and $80 \hmpc$ on a side respectively.  We refer to these two
simulations as ``L120'' and ``L80.''  The corresponding particle
masses in these simulations are $\mpt \simeq 1.07 \times 10^9 \hMsun$
in L120 and $\mpt \simeq 3.16 \times 10^8 \hMsun$ in L80.  Both
simulations use root computational grids of $512^3$ cells and
adaptively refine the grids according to the evolving local density
fields to a maximum of $8$ levels.  This results in peak spatial
resolutions of $h_{\mathrm{peak}} \simeq 1.8 \hkpc$ and
$h_{\mathrm{peak}} \simeq 1.2 \hkpc$ in comoving units for L120 and
L80 respectively.

We identify halos and subhalos (self-bound halos with centers located
within the virial radius of a larger halo) using a variant of the
Bound Density Maxima algorithm \citep[BDM,][]{klypin_etal:99}. Each
halo is associated with a density peak, identified using the density
field smoothed with a 24-particle SPH kernel.  All particles within a
search radius of $\rfind = 25 \hkpc$, set to match the size of the
smallest objects we aim to identify, are removed from further
consideration as potential halo centers.  The BDM algorithm
iteratively removes unbound particles from each halo and uses the
remaining bound particles to calculate halo properties such as the
virial mass $\Mvir$, circular velocity profile $V_{\mathrm{c}}(r) =
\sqrt{GM(<r)/r}$, maximum circular velocity $\Vmax$, and the mass
within a tidal truncation radius.  A more detailed description of the
algorithm and specific parameters used is given in
\citet{kravtsov_etal:04}.

We define a virial radius $\Rvir$, as the radius of the sphere,
centered on the density peak, within which the mean density is
$\Dvir(z)$ times the mean density of the universe, $\rhomean$.  The
virial overdensity $\Dvir(z)$, is given by the spherical top-hat
collapse approximation and we compute it using the fitting function of
\citet{bryan_norman:98}. In the $\Lambda$CDM cosmology that we adopt
for our simulations, $\Dvir(z=0) \simeq 337$ and $\Dvir(z) \rightarrow
178$ at $z \gtrsim 1$.  In what follows, we use virial mass to
characterize the masses of distinct host halos (i.e., halos whose
centers do not lie within the virial radius of a larger system).  We
quantify the sizes of subhalos using their maximum circular
velocities, $\Vmax$, because $\Vmax$ is measured more robustly in
dense environments and, unlike mass, is not subject to the ambiguity
of a particular definition.

\subsection{Correlation Statistics}
\label{sub:mcf}

We quantify the dependence of clustering on halo properties using the
statistics of marked point distributions.  For each halo property, or
{\em mark}, quantified by some value $m$, the distribution of $m$ over
all halos may be characterized by the standard one-point statistics,
the mean $\avg{m}$, the variance $\var{m}$, and higher order moments.
To quantify the dependence of clustering upon $m$ one can construct
the {\em mark-correlation function (MCF)} $k_{\rm mm}(r) \equiv
\pairavg{m_1 m_2}(r)/\avg{m}^2$ , with mark $m$
(\citealt{beisbart_kerscher:00}; see \citealt{gottlober_etal:02} for
another definition; \citealt{sheth:05}).  The notation indicates that
$\pairavg{m_1 m_2}(r)$ is the average of the product of $m_1$ and
$m_2$ for halos at points $\vec{x}_1$ and $\vec{x}_2 = \vec{x}_1 +
\vec{r}$ in pairs separated by distance $r = \vert \vec{r}\vert$.
Similarly, one can compute the average value of $m$, $\pairavg{m}(r)$,
on the condition that a halo is part of a pair at separation $r$.
This formalism can easily be extended to discrete marks, like Hubble
type or number of satellite halos
\citep[e.g.,][]{beisbart_kerscher:00}.

Values of $k_{\rm mm}(r) > 1$ indicate preferred clustering of 
halos with $m$ higher than average.  
The magnitude of deviations of $k_{\rm mm}(r)$ from
$1$ is set by the size of fluctuations of $m$.  In what follows, 
we employ a modified MCF that is normalized 
to the intrinsic one-point fluctuations in $m$, namely 
\begin{equation}
\label{eq:mymcf}
\mymcf{m}(r) \equiv (\pairavg{m_1 m_2}(r) - \avg{m}^2)/\var{m}.
\end{equation}
In the absence of spatial segregation on $m$,
$\pairavg{m_1 m_2}(r) = \avg{m}^2$ and $\mymcf{m}(r)=0$.  
The deviations from the case of 
no segregation are expressed in units of $\var{m}$.  
Roughly speaking, a value of $\mymcf{m}(r)=0.25$ is 
indicative that halos in a pair separated by distance 
$r$ have values of $m$ that are $0.5\sigma$ 
($=\sqrt{0.25}\sigma$) higher than $\avg{m}$.  

\citet{beisbart_kerscher:00}, \citet{gottlober_etal:02}, and
more recently, \citet*{sheth_etal:05}, have
used MCFs to study luminosity- and morphology-dependent clustering in 
observational, simulated, and semi-analytic samples, respectively. 
\citet{sheth_tormen:04} and \citet{harker_etal:06} used the 
statistics of marked point distributions 
to study the dependence of halo clustering on mass assembly
history in cosmological simulations.  In this paper, we apply 
mark-correlation functions to study the environmental 
dependence of halo concentrations and subhalo abundance.

\subsection{Halo Concentrations}
\label{sub:cvir}

The spherically-averaged density profiles of cosmological halos can be
described by the profile of \citet*[][hereafter NFW]{navarro_etal:97}
\begin{equation}
\label{eq:nfw}
\rho(r) = \rho_{0} ({r}/{r_{\mathrm{s}}})^{-1}
(1 + {r}/{r_{\mathrm{s}}})^{-2}.
\end{equation}
The transition radius between the inner and outer power laws 
$r_{\mathrm{s}}$, is often quantified by the concentration 
parameter $\cvir \equiv \Rvir/r_{\mathrm{s}}$.  In the next 
section, we quantify the dependence of halo clustering on $\cvir$.  
We assign each halo a best-fit concentration by fitting 
halo density profiles in logarithmically-spaced radial bins 
following \citet[][B01 hereafter]{bullock_etal:01}.  We consider only halos 
with more than $250$ particles within their virial radii, 
resulting in lower mass limits of 
$M_{\mathrm{min}} \simeq 2.7 \times 10^{11} \hMsun$ 
for L120 and 
$M_{\mathrm{min}} \simeq 7.9 \times 10^{10} \hMsun$ 
for L80.  In addition to the L120 and L80 simulations, 
we re-analyzed the simulation of B01
for the mean $\cvir$-$\Mvir$ relation.  The B01
simulation had the same formal spatial and mass resolution as 
L120 in a computational box of $60 \hmpc$ on a side.  The cosmology 
adopted in B01 was a $\Lambda$CDM cosmology with 
a power spectrum normalization of $\sigma_{8} = 1.0$.  
Hereafter, we refer to this simulation as ``L60.''
We have reanalyzed the L60 simulation and
reproduced the $\cvir(\Mvir)$ results of B01 using our techniques.

The mean relation for $\cvir(\Mvir)$ in the L80 and L120 simulations
is well described by the model of B01 {\em with modified model
  parameters.}  The scatter in $\cvir$ at fixed halo mass is well
described for both simulations by a log-normal distribution with a
standard deviation of $\sigma ( \log \cvir ) \simeq 0.14$ at all
probed masses in accordance with B01 and \citet{wechsler_etal:02}.  In the
notation of B01, we find that our $\cvir(\Mvir)$ relation is well
described by the B01 model with $F=10^{-3}$ and $K=2.9$.  The revised
parameters result in a shallower scaling of concentration with mass,
and slightly lower values of the concentration around $M_*$.  It also
predicts concentration values that are about 20\% lower at $10^{11}
\hMsun$, but note that this is still an extrapolation of the model to
lower mass scales than it has been measured robustly by our
simulations.  A similar revision of the B01 parameters was previously
proposed by \citet[][see also \citealt{kuhlen_etal:05} who advocated
  lower $K$] {dolag_etal:04} to accommodate their simulations of
cluster-sized halos and is in broad agreement with the profiles of
observed clusters \citep{vikhlinin_etal:06}.  The differences in the
$\cvir(\Mvir)$ relation are attributable to a combination of the
differences in the initial power spectra of the simulations and cosmic
variance due to the finite sizes of the computational volumes.

%
%
\begin{figure}[t]
\epsscale{1.1}
\plotone{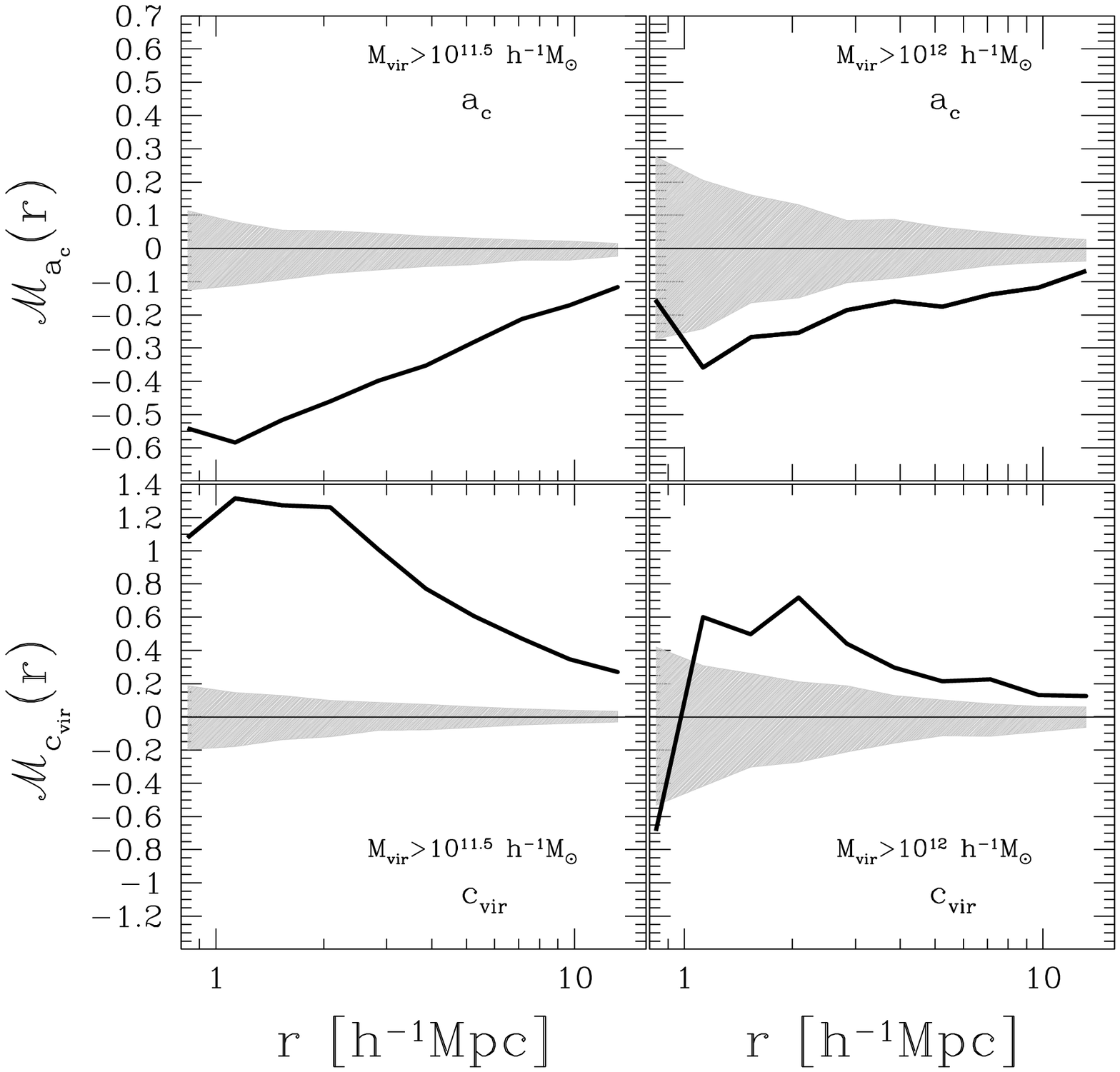}
\caption{
  Clustering and halo properties at $z=0$.  
  The top panels show MCFs 
  $\mymcfa(r)$ ({\em solid lines}), 
  with normalized formation time $\amark$, 
  as mark for two different mass cuts, 
  $\Mvir \ge 10^{11.5} \hMsun$ ($\ge 300$ particles)
  and $\Mvir \ge 10^{12} \hMsun$ ($\ge 940$ particles).
  The {\em shaded bands} represent 
  the 95$^\mathrm{th}$ percentile of $\mymcfa(r)$ formed from 
  $200$ random reassignments of the marks to halos in the sample.  
  The bottom panels show MCFs $\mymcfc(r)$ with 
  normalized concentration $\cmark$ as 
  the mark at the same host halo mass thresholds.  
  The lines and the shaded band have the same significance 
  as the top panels.  
}
\label{fig:mcfcm}
\end{figure}

\subsection{Halo Formation Times}
\label{sub:ac}

For each halo in our sample, we have determined a mass accretion
history $\Mvir(a)$, by identifying the most massive progenitor of each
halo as a function of time using an algorithm similar to that of
\citet[][more details are given in \citeauthor{allgood:05}
  \citeyear{allgood:05}]{wechsler_etal:02}.  \citet{wechsler_etal:02}
found that the halo mass accretion histories can be characterized by a
one-parameter family of trajectories of the form
\beq
M(a) = \Mobs {\rm exp} \left[-2 \ac \left(\frac{\aobs}{a}-1\right)\right],
\label{eq:mah}
\eeq 
where $\aobs$ and $\Mobs$ are the scale factor and mass at the time
the halo is observed.
Equation~(\ref{eq:mah}) defines a formation scale factor 
$\ac$. We assign to each halo a value of $\ac$ according 
to the value that best fits its mass accretion history, 
following \citet{wechsler_etal:02}.  In the following 
section, we address halo clustering as a function of $\ac$.  

The formation time $\ac$ has a number of advantages over other
definitions of halo formation times, such as the times when halos
first acquire fixed fractions of their final mass.  The quantity $\ac$
is less sensitive to individual events in the formation of a halo, as
it is based on the entire mass accretion history of each halo, rather
then a single epoch.  As shown by \citet{wechsler_etal:02}, it also
has the property that its distribution and average value are only a
function of mass, and not redshift.

\citet{wechsler_etal:02} showed that the formation time $\ac$ is
tightly correlated with $\cvir$: the mean relation given by $\cvir =
c_1/\ac$, where $c_1$ is the concentration of halos forming today.  In
the $\Lambda$CDM cosmology adopted in our simulation, these correspond
to halos with $M \sim 10^{15}\hMsun$, and $c_1 \sim 4$ (see
\citealt{wechsler_etal:02} for details).  At fixed redshift,
$\ac(\Mvir)$ is a weak function of $\Mvir$ and the distribution of
$\ac$ at fixed mass can be characterized by a log-normal distribution
with $\sigma(\log \ac) = \sigma(\log \cvir) \simeq 0.14$
\citep{wechsler_etal:02}.

\begin{figure}[t]
\epsscale{1.1}
\plotone{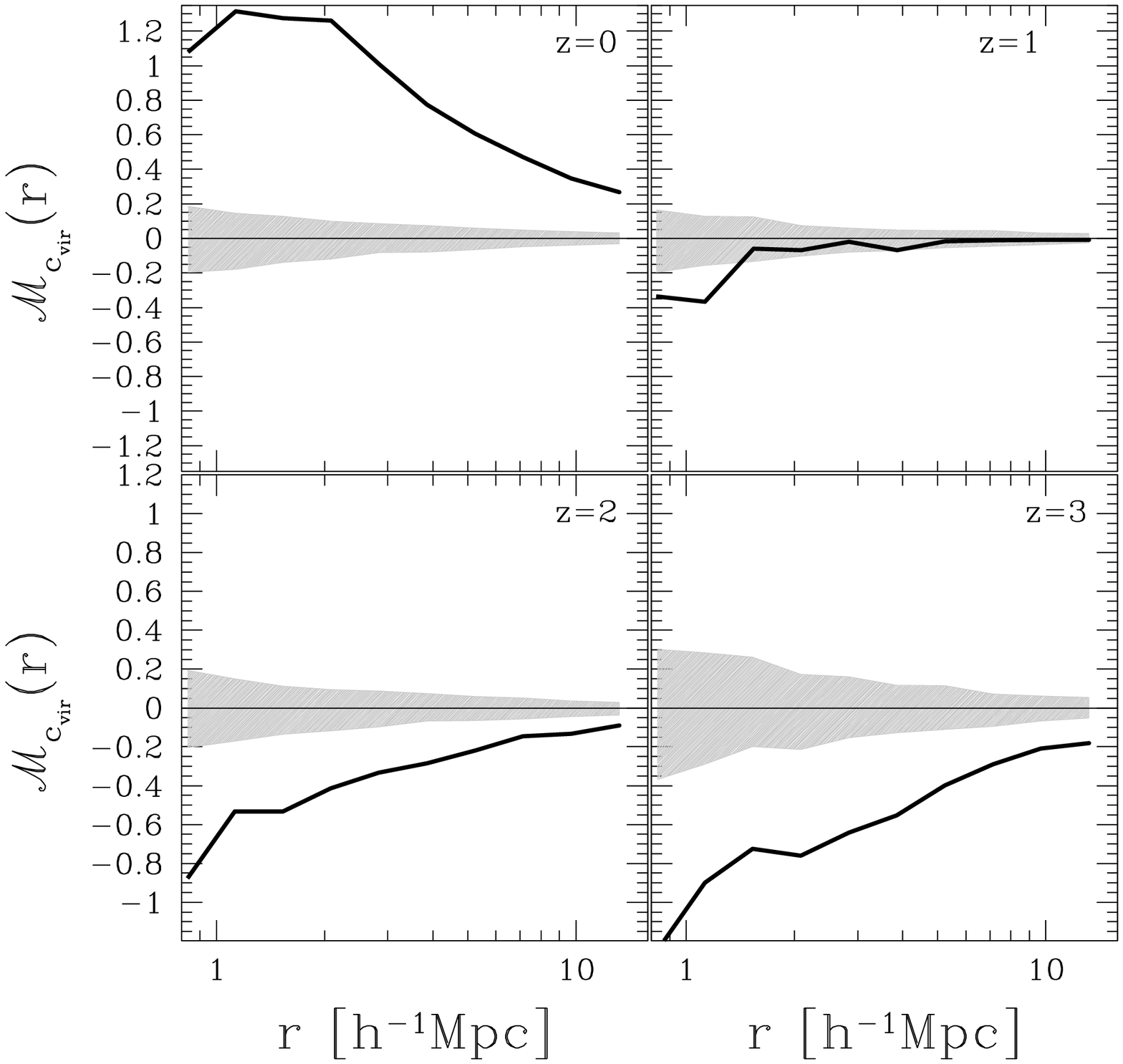}
\caption{
The dependence of clustering on $\cvir$ 
  as a function of redshift.  The panels show MCFs 
  $\mymcfc(r)$ ({\em solid lines}) with normalized
  halo concentration $\cmark$, as mark for halos with 
  $\Mvir \ge 10^{11.50} \hMsun$ at four different 
  redshifts.  {\em Shaded bands} 
  represent the 95$^\mathrm{th}$ percentile of $\mymcfc(r)$ 
  from $200$ random reassignments of 
  the marks.
}
\label{fig:mcfcv}
\end{figure}

\section{Results}
\label{section:results}

\subsection{Formation Time and Concentration Marks}
\label{sub:accmark}

Figure~\ref{fig:mcfcm} shows mark-correlation functions with halo
formation time and concentration used as marks.  The simple mass
dependence of $\ac$ and $c_{\rm vir}$, and the simple distributions of
these quantities at fixed mass allow us to scale out the gross mass
dependence of these quantities in studying formation time- and
concentration-dependent clustering.  We accomplish this by assigning
each halo a normalized formation time $\amark \equiv
\ac/\avg{\ac({\Mvir})}$ and concentration $\cmark \equiv
\cvir/\avg{\cvir(\Mvir)}$, where $\avg{\ac({\Mvir})}$ and
$\avg{\cvir(\Mvir)}$ are the averages of formation time and
concentration as a function $\Mvir$ computed in bins of width $\Delta
\log (\Mvir) = 0.20$.

Consider first the top panels of Figure~\ref{fig:mcfcm}, where we show
MCFs $\mymcfa(r)$ with mark $\amark$, for distinct halos above two
mass thresholds, $\Mvir \ge 10^{11.5} \hMsun$ and $\Mvir \ge 10^{12}
\hMsun$.  We represent the statistical significance of deviations of
the MCF from the null hypotheses of the absence of spatial segregation
on formation time or concentration by randomly reassigning the marks
among the halos in the sample $200$ times and recomputing the MCFs on
these random samples.  The shaded regions in Figure~\ref{fig:mcfcm}
and all MCF figures that follow show the envelope formed by $95\%$ of
these randomized MCFs.

Figure~\ref{fig:mcfcm} shows a statistically-significant tendency for
early-forming (low $\ac$) halos to be more strongly clustered in both
mass bins.  The strength of the trend diminishes with increasing mass.
These results are in qualitative agreement with \citet{gao_etal:05}
and \citet{harker_etal:06}, who define formation time as the time the
halo first acquired half of its final mass.  We find a similar signal
for this definition of formation time as well.  This figure also
clearly shows a dependence with scale, with stronger trends at a few
$\hMpc$ than on larger scales (note that the spread in the range for
randomized samples indicates larger errors at small radii due to fewer
pairs of halos in these bins).  In what follows, we explore the
redshift dependence of clustering as a function of concentration
$\cvir$, because robust determinations of $\ac$ become increasingly
difficult at high redshift as smaller portions of the halo mass
accretion histories are sampled.  

The known relationship between formation time and $\cvir$ suggests
that halo clustering should be a strong function of $\cvir$ unless the
relationship between $\cvir$ and $\ac$ is itself a strong function of
environment.  Neither \citet{lemson_kauffmann:99}, looking for trends
with density, nor \citet{sheth_tormen:04}, using mark-correlation
statistics, were able to detect any significant clustering segregation
with concentration.  However, the simulations employed in these
studies were not particularly well suited to resolve the detailed
density structures of halos, especially at low mass where the trends
are strongest.  Our simulations cover a similar computational volume,
but they have substantially higher mass and force resolutions compared
with these earlier studies.

Figure~\ref{fig:mcfcm} clearly shows a tendency for preferential
clustering of halos selected by their concentrations.  In the bottom
panels, we show $\mymcfc(r)$ at $z=0$ for host halos above two mass
thresholds.  As might have been expected based on the $\ac$-dependent
clustering, halos with high concentrations are more strongly clustered
than average.  As was the case for formation times, the strength of
the $\cvir$-dependent clustering is striking at the lowest masses.
Halos with $\Mvir \ge 10^{11.5} \hMsun$ in pairs separated by $\lsim 3
\hMpc$ tend to have values of $\cvir$ more than $\sim 1 \sigma$ above
the mean relation.  The statistical significance of this preferential
clustering persists to separations $r > 10 \hMpc$, where halos in
pairs have $\cvir$ values $\gsim 0.5\sigma$ above the mean.  Just as
with $\ac$, the $\cvir$-dependent clustering is a decreasing function
of halo mass over this range.  The figure indicates that in our
simulation, the dependence of clustering on concentration is even
stronger than the dependence of clustering on halo formation time.  We
have verified this by remaking the bottom half of
Figure~\ref{fig:mcfcm} with the variable $c_{ac} \equiv c_1/a_c$ as the
mark.  This plot looks quite similar to the inverse of the formation
time mark, and does not show as strong of a signal as the measured
concentration.  It seems likely that this discrepancy is just due to
larger measurement errors in formation time, but a more detailed
analysis with a larger simulation will be necessary to determine this.

We explore $\cvir$-dependent clustering at fixed virial mass ($\Mvir >
10^{11.5} \Msun$) as a function of redshift in Figure~\ref{fig:mcfcv}.
Interestingly, this effect is a strong function of redshift.  Indeed
the sense of the clustering trend reverses over the redshift range
shown here.  Above this fixed absolute mass threshold,
high-concentration halos are more strongly clustered at $z=0$, while
at $z \sim 1$ there is, at most, a weak trend and at $z \gsim 2$ {\em
low-concentration} halos tend to be clustered more strongly.  This
tendency for halos with low $\cvir$ values to be more weakly clustered
at high redshift increases steadily with redshift thereafter.  This
trend suggests that late-forming (high-$\ac$) halos are actually more
strongly clustered at high redshift though we show no direct
statistically-significant evidence of this, largely due to the
difficulty in making robust determinations of $\ac$ at high redshift.

\subsection{Concentration-Dependent Halo Bias}
\label{sub:chalobias}

Due to resolution requirements for measuring $\cvir$, we sample halos
above a fixed mass at each redshift in order to compute the MCFs in
Figure~\ref{fig:mcfcv}, but the typical collapsing mass $\mstar$, is a
declining function of $z$.  At $z=0$, $\mstar \simeq 8.4 \times
10^{12}\hMsun$, while by $z=2$, $\mstar \simeq 1.9 \times
10^{10}\hMsun$, so objects at fixed mass become increasingly rare with
increasing $z$.  It is natural to suspect that the mass and redshift
dependence found in Figures ~\ref{fig:mcfcm} and ~\ref{fig:mcfcv} may
have a common origin, due to $\cvir$- or $\ac$-dependent clustering
that is a function of the relative rarity of the peaks from which
these halos form in the primordial density field
\citep[e.g.,][]{mo_white:96}.  To test this hypothesis, we explore the
relative clustering of halos as a function of concentration and scaled
mass $\Mvir/\mstar$.

\begin{figure}[t]
\epsscale{1.10}
\plotone{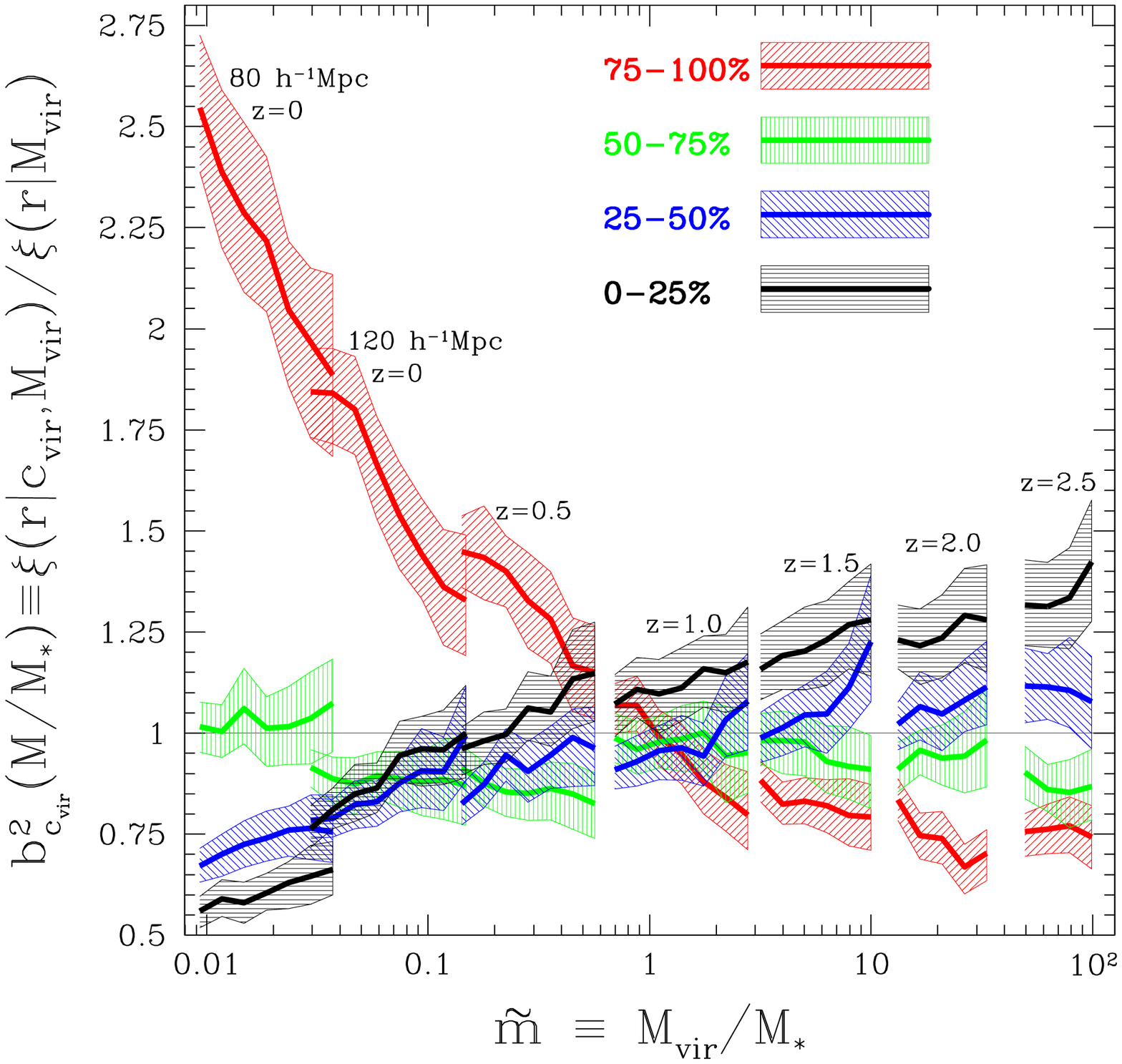}
\caption{
Relative bias squared for halo samples selected by 
quartiles in $\cmark$ and thresholds in the
mass variable $\mmark$,
compared to the bias of all halos above the 
same mass threshold.  Each set of curves shows 
the mean bias for the indicated $\cmark$ quartile.  
The shaded bands represent the $68\%$ region constructed 
from $200$ random subsamples of the unbiased population 
with the same size as the biased subsample.  
The leftmost segments are taken from the $z=0$ 
output of the L80 simulation and are labeled by 
``$80 \hMpc$''.  The remaining segments are taken from 
different redshift outputs of the L120 simulation (labeled ``$120 \hMpc$'') 
as indicated in order to fill in the entire range of $\Mvir/\mstar$.
The left edge of each segment is determined by a minimum of 
$250$ particles in a halo, while the right edge is limited by 
requiring that there be more than $1500$ halos in each subsample.  
}
\label{fig:brel}
\end{figure}

We are unable to test this mass scaling over a large dynamic range at
a single redshift due to the limited dynamic range of our simulations.
In order to explore this $\Mvir/\mstar$ scaling we use the L80
simulation at $z=0$ to explore the low-$\Mvir/\mstar$ regime and we
use several timesteps from the L120 simulation to probe higher values
of $\Mvir/\mstar$.  To quantify $\cvir$-dependent clustering, we
define a {\em relative} bias for a subsample of halos compared to all
halos in the same mass range,

\begin{equation}
\brel^2(r | \mmark) = \xi_{\rm subsample}(r| \mmark)/\xi_{\rm all}(r | \mmark),
\end{equation}

where we have defined a scaled mass variable $\mmark \equiv
\Mvir/\mstar$.  For each subsample, $\brel(r)$ is consistent with
being constant over $5 \le r/\hMpc \le 10$, so we reduce this function
to one number: $\brel^2$ taken over the range of halo separations $r =
5-10\hMpc$.  In Figure~\ref{fig:brel}, we compare this relative bias
as a function of scaled halo mass $\brel^2(\mmark)$ for several
subsamples selected on percentiles of $\cmark.$ The shaded bands
account for the measurement error in $\brel^2$ due to sub-sampling the
full halo distribution by recomputing $\xi_{\rm all}(r)$ from $200$
random subsamples of the total halo population with the same number of
objects as contained in each subsample.  The bands represent the
contours containing $68\%$ of the $\brel^2$ values computed in this
manner.  Different ranges in $\Mvir/\mstar$ are covered by simulation
outputs at different redshifts as labeled in the figure.  Scaling the
results by $\mstar$ delineates a well-defined trend in this
concentration bias as a function of $\mmark$.  In each redshift range,
we are limited at the low-mass end by resolution; we require that a
halo have at least $250$ particles within its virial radius in order
to be considered.  At the high-mass end, the bands are limited by the
requirement that there be at least $1500$ halos in each subsample.
These requirements give rise to the finite length of each segment in
Figure~\ref{fig:brel}.  Results from the L80 simulation at $z \sim 1$
are in good agreement with the L120 simulation at $\mmark \sim 0.1$;
however, we do not plot these in the interest of clarity.

Figure~\ref{fig:brel} clearly demonstrates the trend already indicated
by the mark-correlation functions for the highest-$\cvir$ halos to be
much more strongly clustered than average for $\Mvir \lsim \mstar$ and
less strongly clustered than the overall halo population for $\Mvir
\gsim \mstar$.  It is worth noting at this point that above $\mstar$,
where the scaling of bias with mass is very strong, mass is still the
dominant variable in determining bias.  However, well below $\mstar$,
the scaling of bias with mass flattens, and formation time appears to
be the dominant variable determining bias.

Below, we provide a fitting function for the bias as a function of
both concentration and mass.  Our simulation data are not sufficient
to determine this function with high accuracy, but we present this
function to give a convenient way to estimate the magnitude of the
effects of these trends in particular applications such as the 
clustering of specific galaxy populations.

Let $\cp \equiv \ln(\cmark)/\sigma (\ln \cvir) = \log(\cmark)/\sigma
(\log \cvir)$, such that the probability distribution of $c'$ at fixed
halo mass $P(\cp) {\mathrm d}\cp$ is Gaussian with unit variance.  We
define the relative bias of halos as a function of $\cp$, as the ratio
of the clustering amplitude of halos of fixed $\mmark$ and fixed $\cp$
relative to the clustering amplitude of all halos of fixed $\mmark$,
\begin{equation}
\brel^2(\cp | \mmark) \equiv \frac{\xi(r, \cp |   \mmark)}{\xi(r|\mmark)},
\end{equation}
where here we have again taken the average of the halo bias over 
separations from $5 \le r/\hMpc \le 10$.  
We find that a good fit to the simulation data is given by  
\begin{equation}
\brel(\cp | \mmark) 
= p(\mmark) + q(\mmark) \cp + 1.61[1-p(\mmark)] {\cp}^2,
\label{eq:brel}
\end{equation}
where
\begin{eqnarray}
p(\mmark) & = & 0.95 + 0.042 \ln(\mmark^{0.33}) \nonumber \\
q(\mmark) & = & 0.1 - \frac{0.22[\mmark^{0.33} + \ln(\mmark^{0.33})]}{[1 + \mmark^{0.33}]}. \nonumber
\end{eqnarray}

\begin{figure}[t]
\epsscale{1.10}
\plotone{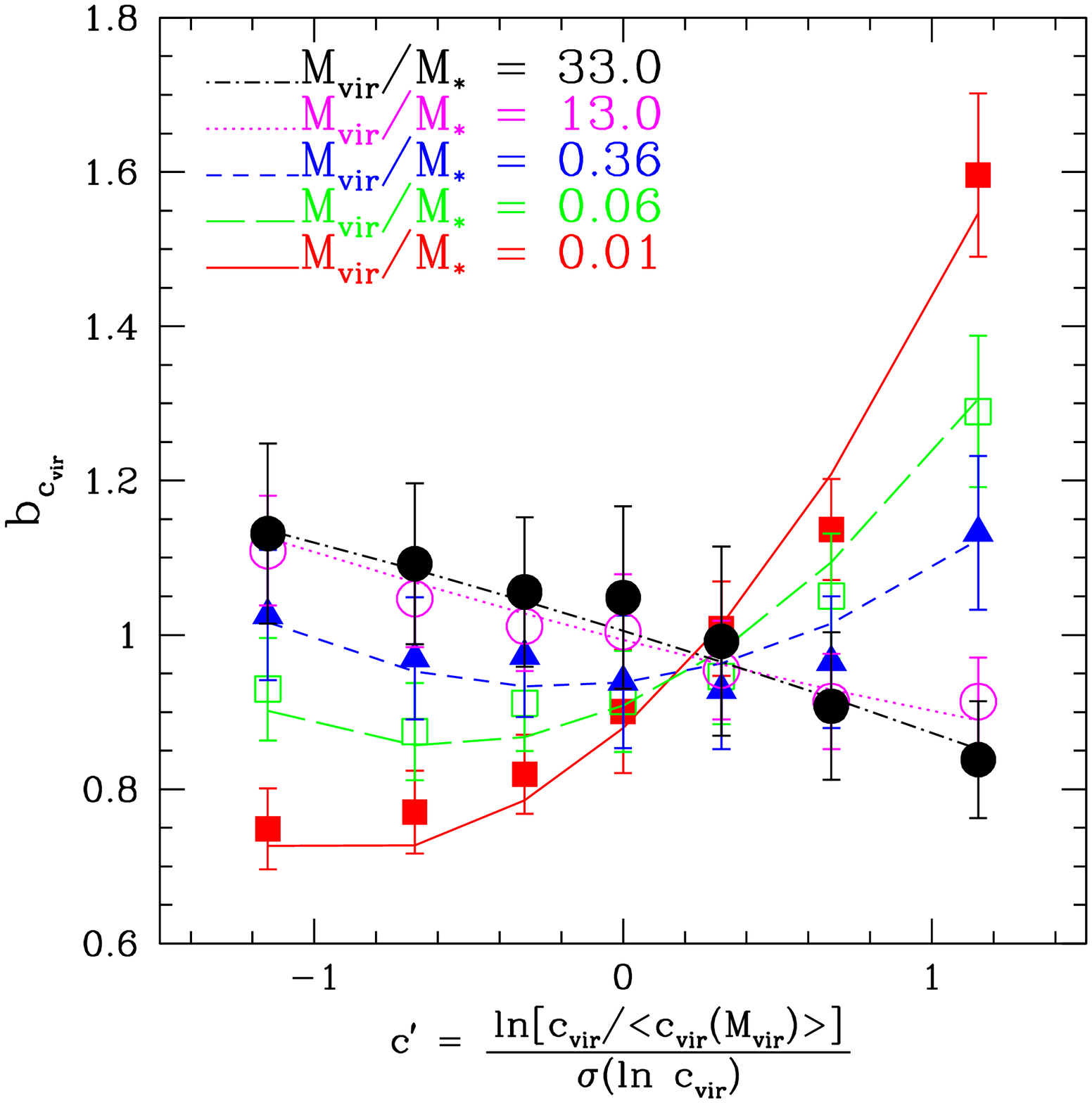}
\caption{
Relative bias as a function of normalized concentration, 
$\cp  \equiv \log( \cmark )/\sigma( \log \cmark )$,
for various values of halo mass scaled by the typical collapsing 
mass $\mmark \equiv \Mvir/\mstar$.  The {\em points} 
show the values of the relative bias, $\brel$, measured 
directly from the host halos in the L120 and L80 simulations 
and the {\em lines} show the fit of Eq.~(\ref{eq:brel}).  
We show the relative bias at five different scaled halo masses, 
$\mmark = 33$, $\mmark = 13$, $\mmark = 0.36$, $\mmark = 0.06$, 
and $\mmark = 0.01$.  
}
\label{fig:bfit}
\end{figure}

The best fitting parameters for this relation satisfy the 
normalization condition
\begin{equation}
\int \brel(\cp | \mmark) P(\cp) {\mathrm d} \cp  = 1.0 
\end{equation}
to within a few percent.  The simulation results are consistent with
this normalization within the sizable errors, and we justify this
constraint in more detail in \S~\ref{section:implications}, where we
discuss the implications of this relative bias of halos on the halo
model.  In Figure~\ref{fig:bfit}, we show the fit of
Eq.~(\ref{eq:brel}) compared to the relative bias measured from the
simulations as a function of the scaled concentration variable $\cp$,
for several values of the scaled halo mass $\mmark$.

\subsection{Halo Occupation Mark}
\label{subsection:hocc}

We expect that this clustering effect may extend to other properties
of halos and the galaxies they host, especially those which are known
to be strongly correlated with formation time and halo structure.
Halo angular momentum and halo shape are two such halo properties that
are relevant to galaxy formation and known to correlate well with halo
formation history \citep[e.g.,][]{vitvitska_etal:02,allgood_etal:06}

The quantity from dissipationless simulations that is most pertinent
to models of the statistics of galaxy clustering is $P(\Nsat|\Mvir)$,
the probability distribution of the number of subhalos per host halo
at fixed host halo mass.  In such simulations this is the best proxy
for the number distribution of satellite galaxies per halo
\citep[e.g.,][]{kravtsov_etal:04}.  The probability distribution of
this number of satellite galaxies per halo as a function of halo mass,
$P(\Nsat|\Mvir)$ is a primary ingredient in halo model calculations of
galaxy clustering (see \S~\ref{section:implications} below).
\citet{sheth_tormen:04} and \citet{gao_etal:05} have emphasized that
the formation time dependence of clustering breaks a fundamental
assumption of the halo model, namely that galaxies populate halos of a
given mass in a manner that is statistically independent of halo
environment.  In fact, this is true only if $P(\Nsat|\Mvir)$ is a
function of halo formation time.  Subhalos are natural sites for
galaxy formation, so a more direct test is to show that halos cluster
differently as a function of $\Nsat$.

\begin{figure}[t]
\epsscale{1.10}
\plotone{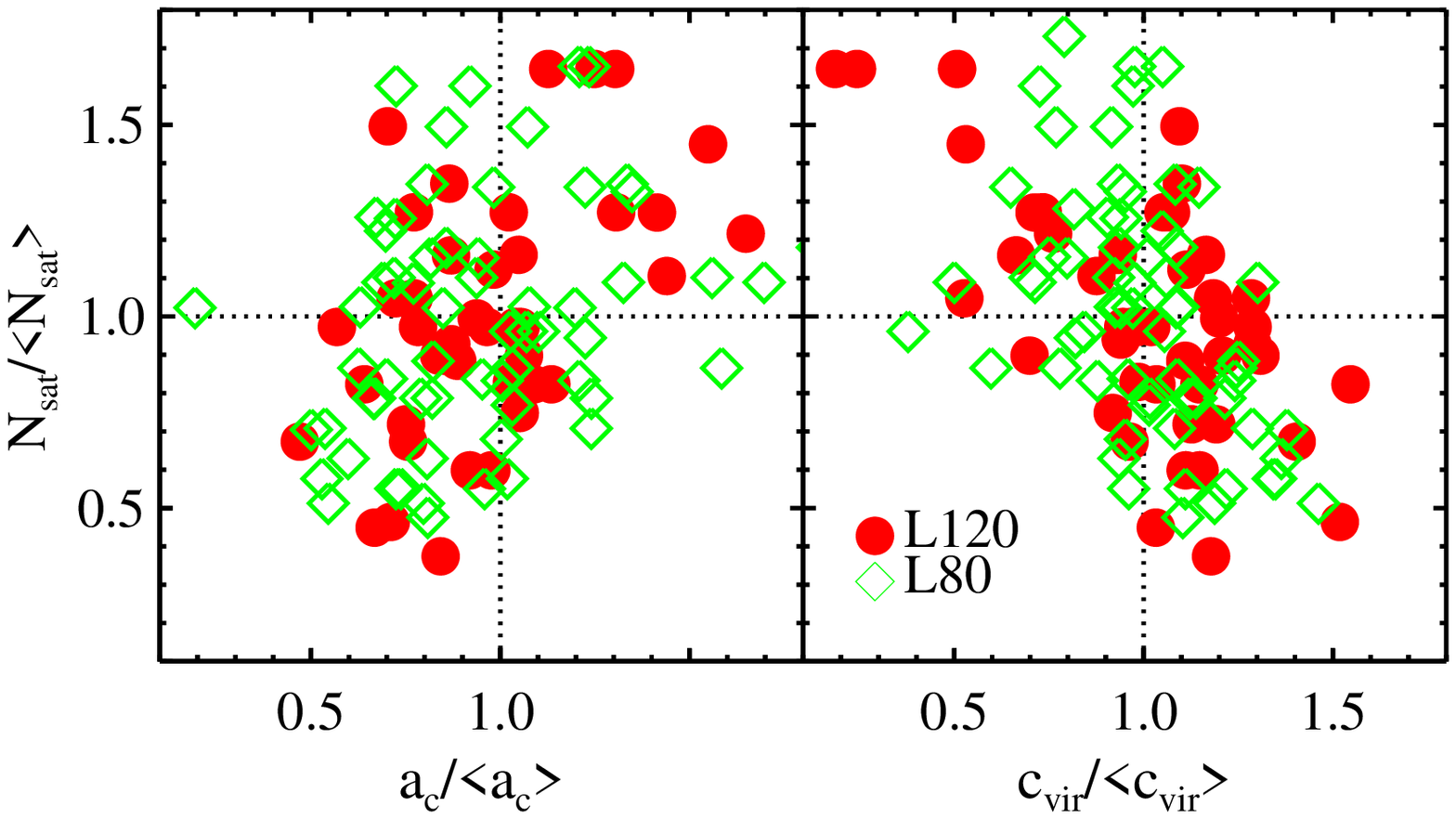}
\caption{Correlation between halo occupation number and formation
  scale factor ({\em left panel}) and concentration ({\em right
      panel}), counting subhalos 1000 times less massive than their
    hosts.  Host halos more massive than $1\times 10^{14}\hMsun$ are
    plotted from the L120 simulation ({\em red circles}), and host
    halos more massive than $5\times 10^{13}\hMsun$ are plotted from
    the L80 simulation ({\em green diamonds}).  Each of the variables
    is normalized to the mean of the variable as a function of halo
    mass.  }
\label{fig:cn}
\vspace{0.1cm}
\end{figure}

\citet{zentner_etal:05} showed that both $\ac$ and $\cvir$ are
strongly correlated with $\Nsat$ in host halos of fixed mass.  We
update this correlation for the massive halos in the L120 and L80
simulations in Figure~\ref{fig:cn}, where we compare the number of
satellites with $M_{\rm host} > 10^{3} M_{\rm sub}$ in the massive
host halos of the L120 and L80 simulations with the host halo
concentrations and formation times.  Scaling the satellite number with
respect to the host mass normalizes out the gross dependence of
satellite number on host halo mass.  Moreover, we have normalized both
$\cvir$ and $\ac$ to their average values as a function of halo mass.
Figure~\ref{fig:cn} clearly shows that early-forming,
high-concentration halos have fewer satellites.  The basic reason is
that halos that accrete their subhalos first have more time for those
subhalos to be destroyed or to merge with the central object due to
dynamical friction \citep[e.g.,][]{kravtsov_klypin:99,taffoni_etal:03,
zentner_bullock:03,zentner_etal:05,vandenbosch_etal:05,taylor_babul:05}.

In light of this strong correlation, the clustering dependence of
formation time and halo concentration found in the previous section
suggests that halo clustering is likely to be a function of $\Nsat$ as
well. \citet{kravtsov_etal:04}, \citet{tasitsiomi_etal:04}, and
\citet*{conroy_etal:06} have demonstrated that halos and subhalos
selected by their maximum circular velocities provide excellent
matches to the observed galaxy-galaxy autocorrelation function,
galaxy-mass cross correlation function, as well as to the
luminosity-dependence and redshift evolution of clustering,
respectively (see also \citealt{berrier_etal:06} for a similar result for
close-pair statistics).  Following these studies and the arguments in
\S~\ref{sub:sim} for quantifying subhalo size as a function of maximum
circular velocity, we study host halo clustering as a function of the
number of satellites above a $\Vmax$ threshold as a quantity that is
particularly relevant to galaxy clustering predictions.

\begin{figure}[t]
\epsscale{1.2}
\plotone{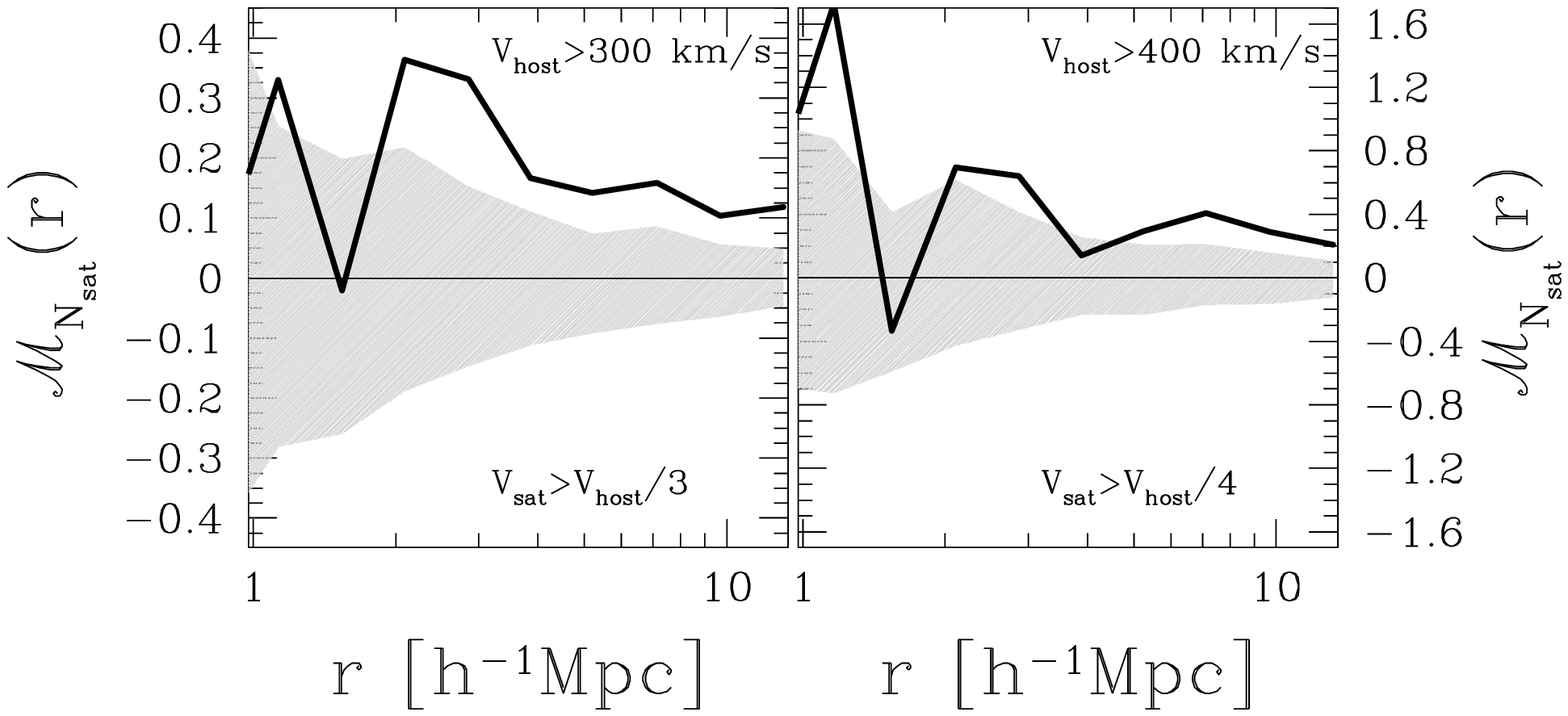}
\caption{
The dependence of clustering on satellite number.
The panels show MCFs $\mymcfN(r)$ ({\em solid lines}), 
with the number of dark matter subhalos $\nmark$, 
as mark.  The two panels correspond to different 
samples.  In the {\em left panel}, results are shown 
for all host halos with maximum circular velocities 
$\Vhost \ge 300\kms$ and 
satellites with 
$\Vsat \ge \Vhost/3$.  In the {\em right panel}, 
we show a sample with $\Vhost \ge 400 \kms$ and 
$\Vsat \ge \Vhost/4$.  
In each panel, the {\em shaded bands} 
have the same meaning as 
in Figure~\ref{fig:mcfcv}.  
}
\label{fig:mcfn}
\end{figure}

In Figure~\ref{fig:mcfn} we make a first attempt to quantify the
strength of clustering for samples of halos marked by their occupation
number.  For two halo samples, we have selected subhalos with maximum
circular velocities $\Vsat$, above a fixed fraction of the maximum
circular velocities of their hosts $\Vhost$.  Taking this ratio of
circular velocities scales out the dependence of $\Nsat$ on host halo
size.  Figure~\ref{fig:mcfn} shows a relatively small but
statistically-significant tendency for halos in pairs separated by
$\sim 5-10 \hMpc$ to have above-average numbers of satellites.  This
is the most direct demonstration yet that the halo occupation by
galaxies is a function of environment.

Note that in Figure ~\ref{fig:mcfn}, we are forced to study only large
host halos in order to guarantee that their subhalos are well
resolved.  As such, we probe a different range of host halo masses
than shown in Figure~\ref{fig:mcfcv}.  The sense of the trend is what
we would expect at this mass range, which is slightly bigger than
$\mstar$, from the correlation between $a_c$ and $\Nsat$.
Low-concentration, late-forming halos in this mass range are more
clustered than average, and it is these halos that are expected to
have more satellites.  Based on the previous results for $\ac$- and
$\cvir$-dependent clustering, if one were able to measure the
clustering of low-mass halos with several satellites, this trend may
reverse.

\section{Property-Dependent Halo Clustering and the Halo Model}
\label{section:implications}

The results of the previous two sections have potentially 
important implications for the halo model of clustering.  
The basic idea behind the halo model framework has a long history,
initially in analytic models that described galaxy clustering as a
superposition of randomly-distributed clusters with specified 
profiles and a range of cluster masses 
\citep{neyman_scott:52,mcclelland_silk:77,peebles:74}.  
The explosion of recent activity in this field has
been partly fueled by the recognition that a combination of this
approach with recently developed tools for predicting the 
spatial clustering of dark matter halos 
\citep[e.g.,][]{mo_white:96,sheth_tormen:99,sheth_etal:01,
seljak_warren:04,tinker_etal:05} 
provides a powerful formalism for analytic calculations 
of dark matter clustering, which can be extended naturally 
to biased galaxy populations. 

In modern implementations of the halo model
\citep[e.g.,][]{scherrer_bertschinger:91,seljak:00,ma_fry:00,
  peacock_smith:00,scoccimarro_etal:01} two-point galaxy clustering is
calculated by specifying the clustering of dark matter, the non-linear
clustering of dark matter halos, the first two moments of the HOD, and
the spatial distribution of galaxies within their host halos.  The
standard implementation also assumes that {\em halo clustering is
  independent of all halo properties aside from halo mass.}  In
particular, in calculations of either galaxy or dark matter
correlation functions, it is assumed that the HOD and halo
concentrations depend only on halo mass and that there is no spatial
segregation of halos based on their occupation numbers,
concentrations, or any other properties that may be relevant to the
properties of the galaxies that the halos host.  We refer to a halo
model based on this set of assumptions as the ``strong'' halo model.
As we have shown, these assumptions are not generically valid: the
two-point clustering of dark matter halos depends on halo
concentration, halo formation time, and halo occupation.  Below, we
review several salient aspects of the halo model and explore the
implications of relaxing these assumptions about the lack of
environmental dependence of halo properties.  As an illustrative
example that is closely tied to the results of the previous section,
we focus most of our attention on relaxing the assumption that
clustering is independent of the halo profile concentration.  It is
worth keeping in mind that there are several different implementations
of various aspects of the halo model: choices must be made about how
to model the translinear regime for galaxies (especially the treatment
of halo exclusion and scale-dependent bias, see
e.g. \citealt{tinker_etal:06}, Appendix B), what analytic models to
use for the mass function and bias, and how to model the halo occupation
\citep[e.g.]{zheng:04, conroy_etal:06} and the profiles of galaxies in halos.

\subsection{The Dark Matter Correlation Function in the Standard Halo Model}
\label{sub:dmstandard}

The simplest application of the halo model is to calculate the dark
matter two-point correlation function.  The halo model breaks the computation
into a ``one-halo'' term receiving contributions from the mass density
in individual halos on small scales and a ``two-halo'' term with
contributions from mass in distinct pairs of halos on large scales.
In the standard halo model, we can write the large-scale halo
correlation function of dark matter as
\citep{scherrer_bertschinger:91}:
\begin{equation}
\label{eq:decomp}
\xidm(r) = \xioh(r) + \xith(r) + 1.
\end{equation}
The one-halo term is 
\begin{equation}
\label{eq:xi1h}
\xioh(r) = \frac{1}{\rhomean^2}
\int \dd m\, m^2 \frac{\dd n(m)}{\dd m} 
\int \dd^3 x\, \lambda_m (\vec{x}) \lambda_m(\vec{x}+\vec{r}), 
\end{equation}
with $\dd n(m)/\dd m$ the mass function of halos and 
$\lambda_{m}(\vec{x})$ the density distribution 
within a halo of mass $m$ normalized so that the integral 
of the profile over the volume of the halo is unity.  The 
two-halo term is 
\begin{eqnarray}
\label{eq:xi2h}
\xith(r) & = & \frac{1}{\rhomean^2} 
\int \dd m_1\, \int \dd m_2\, 
m_1 \frac{\dd n(m_1)}{\dd m_1}\, m_2 \frac{\dd n(m_2)}{\dd m_2} \nonumber \\
 & \times & \int \dd^3 x\, \int \dd^3 y\,\,  
\lambda_{m_1}(\vec{x}) \lambda_{m_2}(\vec{y}) \nonumber \\
 & \times & \, \xihh( \vec{x}-\vec{y}+\vec{r}|m_1,m_2), 
\end{eqnarray}
where $\xihh(\vec{x}|m_1,m_2)$ is the cross-correlation function 
of halos of mass $m_1$ and $m_2$ and 
$r \equiv \vert \vec{r} \vert$.

In the limit of separations much larger than the sizes of 
the largest halos, the correlation function is determined 
by the two-halo term alone.  On such large scales, the 
correlation functions vary little over the length scales 
of halos so that 
$\xihh(\vec{x}-\vec{y}+\vec{r}|m_1,m_2) \simeq \xihh(\vec{r}|m_1,m_2)$,
which allows the last two integrals in Eq.~(\ref{eq:xi2h}) to be
replaced by $\xihh(\vec{r}|m_1,m_2)$.  Relating the halo 
correlation functions to the dark matter correlation function 
through the standard assumption 
$\xihh(r|m_1,m_2) \simeq \bh(m_1)\bh(m_2)\xidm(r)$
and requiring $\xith(r) = \xidm(r)$ on large scales
forces the halo bias $\bh(m)$ to obey the constraint 
\begin{equation}
\label{eq:bias}
\int \dd m\, \frac{\dd n(m)}{\dd m} \Bigg(\frac{m}{\rhomean}\Bigg) \bh(m) = 1.
\end{equation}
This is the well-known normalization rule for the 
mass-dependent halo bias.  

\subsection{The Galaxy Correlation Function in the Halo Model}
\label{sub:galstandard}

This model can be used to compute the statistics of 
any population for which all members reside in dark matter 
halos.  The most popular application is to compute the correlation 
statistics of galaxies.  The equations of \S~\ref{sub:dmstandard} 
can be adapted to this application simply by making the following 
substitutions.  First, take $\rhomean^2 \rightarrow \ngal/2$, where 
$\ngal$ is the mean number density of galaxies.  Take 
$m \rightarrow \langle \Ngal \rangle_{m}$ so that the mean number 
of galaxies per halo of mass $m$ is counted rather than the mass 
per halo.  Take $m^2 \rightarrow \langle \Ngal (\Ngal - 1) \rangle_{m}/2$
so that pairs of galaxies within halos of mass $m$ are counted.  
Finally, take $\lambda_{m}(\vec{x}) \rightarrow \lambda_{m}^{g}(\vec{x})$ 
to represent the the mean distribution of 
galaxies within host halos of mass $m$ rather than the distribution of 
mass within halos.  

Following the logic of the previous section, we can compute 
the large-scale clustering of galaxies from the two-halo term 
alone.  This leads to the well-known and useful 
relation for the large-scale bias of 
a galaxy population given the first moment of its 
HOD, $\langle \Ngal \rangle_{m}$, 
\begin{equation}
\label{eq:galbias}
b_{\mathrm{gal}} \simeq 
\frac{1}{\ngal} \int \dd m\, \frac{\dd n(m)}{\dd m} 
\langle \Ngal \rangle_{m} \bh(m)\, .
\end{equation}
As one might expect, the large-scale bias of galaxies is 
given by a simple, weighted average of the bias of 
the halos in which they reside.

\subsection{The Dark Matter Correlation Function with $\cvir$-Dependent Halo Clustering}
\label{section:cvirbias2}

Now consider recasting the halo model allowing halo clustering 
to be a function of both halo mass $m$, and the additional 
property of concentration $c$.  
If we define the probability of a value of $c$ 
at fixed mass as $P(c|m)$ then the number of halos 
of mass $m$ with concentration $c$ is 
$\dd n(m,c)/\dd m \dd c = P(c|m)\dd n(m)/\dd m$.   
This additional property complicates the halo 
model and requires an extra integral over the 
distribution of halo concentrations.  Specifically, 
the one- and two-halo terms become
\begin{eqnarray}
\label{eq:xi1hc}
\xioh(r) & = & \frac{1}{\rhomean^2} 
\int \dd m\, \int \dd c\, m^2 P(c|m)\frac{\dd n(m)}{\dd m} \nonumber \\
 & \times & \int \dd^3 x \lambda_{m}(\vec{x}|c) \lambda_{m}(\vec{x}+\vec{r}|c)
\end{eqnarray}
and
\begin{eqnarray}
\label{eq:xi2hc}
\xith(r) & = & \frac{1}{\rhomean^2}
\int \dd m_1 \int \dd c_1 \int \dd m_2 \int \dd c_2 \nonumber \\
 & \times & m_1 \frac{\dd n(m_1)}{\dd m_1} P(c_1|m_1)\, \,
             m_2 \frac{\dd n(m_2)}{\dd m_2} P(c_2|m_2) \nonumber \\
 & \times & \int \dd^3 x\, \int \dd^3 y 
  \lambda_{m_1}(\vec{x}|c_1)\, \lambda_{m_2}(\vec{y}|c_2) \nonumber \\
 & \times & \xihh(\vec{x}-\vec{y}+\vec{r}|m_1,c_1,m_2,c_2)\, ,
\end{eqnarray}
where $\lambda_{m}(\vec{x}|c)$ is the density distribution for a halo of 
mass $m$ and concentration $c$, and where the cross correlation of halos of 
mass $m_1$ and concentration $c_1$ with halos of mass $m_2$ and 
concentration $c_2$ is embodied in 
$\xihh(\vec{x}|m_1,c_1,m_2,c_2)$.  

Two consequences of these relations are evident.  The first is simply
that the one-halo term at small separations should be a weighted
average of profile convolutions with the weighting given by $P(c|m)\,
\dd n(m)/\dd m$.  This is true independent of concentration-dependent
clustering has been overlooked in most modeling efforts (although, 
see e.g. \citealt{sheth_etal:01b}).  The effect of adding scatter is only at
the few percent level, but this will be important for future precision measurements
of dark energy using lensing as the statistical uncertainty of the experiments approaches
this level.  Second, at intermediate scales
($r \sim$ several Mpc), neglecting concentration-dependent clustering
will lead to differences in the calculation of the dark matter
correlation function due to a combination of the preferential
clustering and the convolution factors.

In \S~\ref{sub:dmstandard}, we derived the normalization 
relation for the standard mass-dependent halo bias and an 
analogous relation holds for the concentration-dependent 
relative halo bias $\brel$ that we use in this paper.  We can 
write the halo-halo cross correlation factor in terms of a 
concentration-dependent relative bias of halos with respect to 
all halos at fixed mass defined so that 
$\xihh(r|m_1,c_1,m_2,c_2) = \brel^2(c_1,c_2,|m_1,m_2) \xihh(r|m_1,m_2)$.  
Counting pairs of all halos of fixed mass must give the same result 
regardless of whether or not we subdivide the halo population by 
concentration at fixed mass.  This requirement gives the general 
normalization condition 
\begin{equation}
\label{eq:fullnorm}
\int \dd c_1\, \int \dd c_2\, P(c_1|m_1)\, P(c_2|m_2)\, 
\brel^2(c_1,c_2|m_1,m_2) = 1\, .
\end{equation}
Assuming that we can write the bias term as 
$\brel^2(c_1,c_2|m_1,m_2) = \brel(c_1|m_1) \brel(c_2|m_2)$ 
as in models of deterministic bias, 
the normalization condition is then 
\begin{equation}
\label{eq:4}
\int \dd c P(c|m) \brel(c|m) = 1\, .
\end{equation}
This normalization is consistent with our measurements of the relative
concentration-dependent bias, and holds to within a few percent for
the fitting formulae of \S~\ref{sub:chalobias}.

\subsection{Galaxy Clustering with Concentration-Dependent Bias}
\label{sub:chmgal}

As with the standard halo model, a halo model incorporating
concentration-dependent halo clustering can be formally extended to
galaxy clustering in a simple manner, though this extension may become
cumbersome in practice.  As a first attempt, we may assume that the
number of galaxies per halo is independent of halo concentration so
that as in the previous discussion the substitutions $m \rightarrow
\langle \Ngal \rangle_m$ and $m^2 \rightarrow \langle
\Ngal(\Ngal-1)\rangle_m/2$ can be taken in Eq.~(\ref{eq:xi2hc}) and
Eq.~(\ref{eq:xi1hc}).  The consequences of the concentration-dependent
clustering are then quite similar to the case of the dark matter
correlation function, but are less direct because of inherent
uncertainties in the link between the matter distribution within
halos, $\lambda_{m}(\vec{x}|c)$, and the distribution of galaxies
within halos, $\lambda_m^g(\vec{x}|c)$ \citep{nagai_kravtsov:05,chen_etal:05}.

Another possibility is that the galaxy HOD does depend on host halo
concentration.  Indeed, in Figure \ref{fig:cn}, we confirm that the
number of {\em dark matter} subhalos per host halo at a fixed host
halo mass does correlate with the halo concentration, and we have
shown that host halo clustering is a function of the number of
satellite halos that they contain.  A more general and realistic
assumption then seems to be to consider galaxy number as a function of
both $c$ and $m$ so that the appropriate substitutions into the dark
matter correlation function equations [Eq.~(\ref{eq:xi1hc}) and
  Eq.~(\ref{eq:xi2hc})] are $m \rightarrow \langle \Ngal(m,c)
\rangle_{m,c}$ and $m^2 \rightarrow \langle
\Ngal[\Ngal-1](m,c)\rangle_{m,c}$.  In this case, the large-scale bias
of a particular sample of galaxies can then be written as
\begin{eqnarray}
\label{eq:bgc}
b_{\mathrm{gal}} & \simeq & \frac{1}{\ngal}
\int \dd m\, \frac{\dd n(m)}{\dd m} \bh(m) \nonumber \\
 & \times & \int \dd c\, 
\langle \Ngal(m,c) \rangle \brel(c|m) P(c|m)\, .
\end{eqnarray}
The bias is weighted over  both the concentration 
distribution and the mass distribution.  As such, the 
model can be compared to the standard halo model using 
an effective halo occupation, the mean of which is given by 
\begin{equation}
\label{eq:hodeff}
\tilde{N}_{\mathrm{gal}}(m) = \int \dd c\, P(c|m) \brel(c|m) 
\langle \Ngal(m,c)\rangle \, .
\end{equation}
Note that because early-forming halos have a lower average satellite
number, for luminosity-selected samples it is possible that this
could offset the higher bias and result in {\em galaxy} clustering
that does not depend strongly on $\cvir$; but this is unlikely to be
the case for samples that are more directly connected to formation
time.

These arguments can be used and extended to account for any additional
dependence of halo clustering on halo properties and corresponding
halo occupation distribution, most naturally formation time.  Although
a larger halo sample will be needed to fully characterize these
trends, this first indication of the mass and redshift scaling of the
trends of bias with halo properties that we give here should prove
useful in order to estimate the size of these effects.

\subsection{General Implications}

Although these results urge caution in using the standard halo model
assumptions to calculate galaxy clustering statistics and infer
cosmological parameters with high precision, it is not clear that they
will have a large effect for galaxy samples that are selected by mass
or, as in observational samples, by luminosity.  One indication was
given by the following test.  \citet{zentner_etal:05} used the
standard halo model combined with halo occupation derived from an
analytic model for the evolution of halo substructure to predict the
two-point correlation function for galaxies, and obtained a result
that was consistent with the results of simulations that include all
of the effects mentioned here.  Still, it may be that small
discrepancies would manifest if a larger sample were used to make this
comparison.  Second, \citet{conroy_etal:06} have compared estimates of
the HOD from galaxy clustering measured in the Sloan Digital Sky
Survey (SDSS) \citep{tinker_etal:05}, where these effects are ignored, with
the HOD of subhalos measured in these same simulations, where again
these effects are included implicitly, and found that they are in better than
$\sim 10\%$ agreement for galaxies brighter than $M_r = -19$.

Other authors have investigated whether the halo occupation is
dependent on environment without finding any such trends.  For
example, \citet{berlind_etal:03} investigated mean halo occupation as
a function of local density in hydrodynamic simulation in a
computational box $50 \hMpc$ on a side.  These authors measured local
density in $4 \hmpc$ spheres around host halos and found no indication
of such a trend.  \citet{yoo_etal:05} used the same hydrodynamic
simulation as \citet{berlind_etal:03}, to determine whether
environment-dependent halo occupation effects could be detected in
galaxy correlation statistics.  By swapping the galaxy populations
between halo populations with similar masses, \citet{yoo_etal:05}
found $5-10\%$ effects on the galaxy-galaxy and galaxy-mass
correlations, which were within their statistical uncertainties.

These tests all indicate that the effects for mass- or
luminosity-selected samples are at the $10\%$ or lower level.
However, we caution that these studies all employed simulations of
relatively small volumes.  We have performed a comparable test to that
shown in Figure~\ref{fig:mcfn} using our smaller $60 \hMpc$ box (L60
studied by B01) and found no statistically-significant effect due to
the eightfold smaller computational volume.  As such, it is not
surprising that these earlier studies were unable to find a conclusive
result with their simulation of a yet smaller volume.  This indicates
that large, high-resolution simulations (suitable for detecting
subhalos in $\Mvir < \mstar$ hosts) will be necessary to determine
whether the relative bias measured by halo occupation has the same
behavior as might be expected from using concentration in addition to
the global relationship between concentration and satellite number as
a proxy for halo occupation for all halos in the volume.

These trends have not been readily apparent in observations to date,
although a detailed comparison is complicated by the
difficulties in selecting a sample that closely corresponds to a mass-
and concentration or formation history-selected sample.  Recent studies
looking at galaxy clustering in the SDSS have not found any trend of
clustering with galaxy properties.  \citet{skibba_etal:06} used
luminosity-marked correlation functions to test whether the
luminosity-dependent clustering is consistent with being a simple
consequence of mass-dependent clustering, and found that it is, using
$\sim 0.5L*$ galaxies.  \citet{abbas_sheth:06} showed that the
observed environmental dependence of clustering in SDSS could be fully
accounted for by correlations between galaxy properties and host mass
and by host mass and environment.  Another study by Blanton and
Berlind (in preparation) shuffled $\gsim 0.3 L*$ galaxies between
groups of the same luminosity and found that the red and blue
correlation functions remain unchanged at the $\sim 5\%$ level.

These results offer some confirmation of the theoretical indications
that the effects aren't strong in this regime, but are not terribly
surprising in light of our results, because for halos with
concentrations more than 1$\sigma$ from the mean, the effect is less
than $20\%$ in the range $0.5 \mstar < \Mvir < 10 \mstar$.  Still, the
last study did look at halos well below $\mstar$, and still did not see
a strong trend; this may indicate that the correlations between
formation time and galaxy observables are not extremely strong.  An
additional study was performed by \citet*{yang_etal:06}, who measured
the clustering of a sample of groups selected from the 2dF.  They
investigated the relative clustering of groups of a given mass as a
function of their spectral type, which one might expect to correlate
with formation epoch for galaxies at fixed mass, and found that
clusters with early-type central galaxies were more clustered than
clusters of the same luminosity that had late-type central galaxies.
This goes in the sense of our predicted trend for low mass halos but
they found that it extended to cluster masses.

A first attempt to connect these trends directly with observable
quantities was made by \citet{croton_etal:06} after this paper was
submitted.  This paper uses a semi-analytic galaxy formation model to
investigate the affect of assembly bias on the total galaxy sample and
on samples selected by color, and finds that effects ranging from a
few percent to almost a factor of two depending on the selection.  At
first glance, the predictions of this model do seem to be in mild
conflict with the lack of trends seen in the studies above, but this
is still unclear.  First, shuffling galaxies on observational proxies
for mass may lead to much weaker effects than shuffling on the
theoretical halo mass.  Secondly, the detailed predictions for
observed galaxies may dependent sensitively on the galaxy formation
model, and it may be that the lack of observed trends indicates that
color is less correlated with formation time than in this particular
model.

The trends we have presented here are likely to be more important for
studies of the clustering of extreme objects that may be thought to
form particularly early or late or have particular formation
histories.  Examples of such populations would be color or star
formation selected samples, high-mass clusters selected by occupation
number, low surface-brightness galaxies, or dwarf galaxies.  We
discuss the consequences for such samples further in the next
section.

\section{Summary and Discussion}
\label{section:disc}

We have investigated the clustering of dark matter halos as a function
of several internal halo properties, namely formation time,
concentration, and occupation number.  We have confirmed that halo
clustering is a function of halo formation time and have shown
that the effect is scale-dependent.  We have also
demonstrated that halo clustering is a strong function of halo
concentration, and that the strength {\em and sign} of this trend is a
function of mass.  Of relevance to studies of galaxy clustering
statistics, we also find the clearest indication yet that host halos
of fixed mass cluster in a way that is dependent upon the number of
subhalos that reside in them.  Our primary results can be summarized
as follows.
\begin{enumerate}

\item Halo clustering is a strong function of formation time for fixed
  mass halos.  This effect strengthens with decreasing halo mass
  and with decreasing separations, and is an increasingly strong
  function of mass as halos become less massive than $\mstar$.  These
  results are in broad agreement with the recent results of
  \citet{sheth_tormen:04}, \citet{gao_etal:05} and
  \citet{harker_etal:06}.

\item We have presented the first definitive measurement showing that
  the clustering of dark matter halos is a function of halo
  concentration.  This effect is a strong function of halo mass, and
  can be characterized over a range of mass and redshift as a function
  of halo mass scaled by the typical collapsing mass, $\Mvir/\mstar$.
  Below $\Mvir/\mstar$, halos of high concentration are more clustered
  than halos of low concentration and this trend strengthens with
  decreasing halo mass.  For halos more massive than $\mstar$, the
  trend changes sign, and halos of low concentration become more
  strongly clustered than their high concentration counterparts.

\item The dependence of halo bias on concentration, mass, and redshift
  can be parameterized in a simple way: $b(M,c|z) = b(M|z) \brel
  (c|M/M_*)$.  We provide a fitting function for $\brel(c|M/M_*)$ that 
  can be used to estimate the importance of
  these effects in various regimes.  In \S~\ref{section:implications}, 
  we demonstrate how this relative bias can be incorporated into a 
  halo model formalism and discuss the effects of concentration- and 
  formation time-dependent bias on estimates of matter and galaxy 
  correlation statistics.

\item 
  We confirm and update earlier results 
  \citep[e.g.,][]{gao_etal:04,zentner_etal:05,vandenbosch_etal:05,taylor_babul:05} 
  that the occupation number of satellite halos is strongly 
  correlated with both concentration and formation time.
  We present the first detection of a trend between clustering and
  halo occupation, showing that at high mass, high-occupation (late-forming)
  halos are more clustered than their low-$N$ counterparts.
\end{enumerate}

\citet{sheth_tormen:04}, \citet{gao_etal:05} and
\citet{harker_etal:06} have emphasized that the trend with formation
time indicates that using the halo model to estimate clustering can be
problematic.  We have quantified this explicitly, by investigating how
these trends extend to two variables that are directly relevant to
such calculations, namely concentration and halo occupation.  There is
a weak indication that the trends we see with concentration and halo
occupation are slightly stronger than would be predicted simply by
their dependence on formation time and the trends with formation time
itself, however, this may be do to the larger measurement error in
formation time.  Larger simulations will be required to determine to
higher accuracy whether the strength and nature of the trends with
concentration and occupation have features that are not represented by
the global correlations between these variables and formation time.

As we emphasized in \S~\ref{section:implications}, the dependence of
clustering on concentration implies corrections to halo model
calculations of the {\em dark matter} power spectrum, as well as
corrections to halo model calculations of galaxy correlations.  These
results prescribe caution when using the standard halo model
assumptions to calculate galaxy clustering statistics and infer
cosmological parameters precisely, but it is not clear that they will
have a large effect for samples that are selected by mass or, as in
observational samples, by luminosity.  This is especially true for
galaxies around $L_*$.  For example, \citet{yoo_etal:05} and
\citet{zentner_etal:05} have both shown that shuffling the host halos
of such a sample results in less than about $5 \%$ effects in
clustering statistics, albeit in relatively small volumes.  We expect
that the trends we have shown here will have stronger effects on high-
or low-mass samples (when compared to $\mstar$) that are selected on
some property that is directly connected to either formation time,
concentration, or the number of satellite galaxies, and we speculate
on a few of these below.

The fact that the clustering of low-mass halos is strongly correlated
with formation time may have interesting implications for the
so-called ``void phenomenon'', the tendency of low-mass galaxies to
avoid the voids defined by larger galaxies.  The currently-favored
\LCDM\ model predicts substantial {\em mass} in voids, so the absence
of galaxies in these regions has been suggested to be a potentially
serious problem for the prevailing paradigm \citep{peebles:01}.
Whether this is indeed a problem for these models isn't clear; for
example, \citet{mathis_white:02} have investigated the clustering of
low luminosity galaxies in LCDM simulations combined with a galaxy
formation model and found that all galaxies avoided the voids defined
by the brighter galaxies.  \citet{benson_etal:03} found a similar
result but emphasized that more data was needed to fully test the
issue.  Recently, \citet{furlanetto_piran:06} compared predictions for
void sizes based on the excursion-set formalism with observations from
SDSS, and saw indications that the observed voids are somewhat bigger
than the model predicts.  A related piece of evidence for differential
clustering of low-luminosity galaxies has been emphasized by
\citet{tully_etal:02}, namely, the difference between the luminosity
function in clusters, which rises steeply toward low-luminosity dwarf
galaxies, and the luminosity function in voids, which is substantially
shallower and appears to be entirely bereft of a dwarf galaxy
population.

It has been suggested \citep*[][]{bullock_etal:00} that the
discrepancy between the abundance of dwarf satellites observed in the
Local Group and the abundance of relevant dwarf mass dark halos
($v_{\rm max} \lesssim 50\ {\rm km\, s^{-1}}$) expected by CDM can be
resolved by suppressing galaxy formation in halos that form after the
universe is re-ionized.  This would bias luminous dwarf galaxies to be
associated with late-forming, small dark matter halos.  As we have
shown, the clustering trend with formation time is quite strong in
low-mass halos of this type.  If the photoionization significantly
affects galaxy formation in dwarf-size galaxy halos, the dwarf
galaxies they host would be substantially more clustered than typical
halos in the same mass range, and this trend can be enhanced by
additional environmental factors \citep*{kravtsov_etal:04b}. If we
extrapolate the results of Eq.~(\ref{eq:brel}) to $M \sim 10^{-3}\,
\mstar$, this implies that the correlation length for the earliest
forming quartile should be about a factor of $4$ higher than that of
the typical halos of that mass.  This may provide a natural
explanation for the lack of observed dwarf galaxies in voids.  

More generally, any astrophysical effect that biases small galaxies to
lie in early-forming halos would produce the same effect.  There is
observational evidence that star formation timescales are quite long
($\sim 10$Gyr) in small galaxies and relatively short ($\sim 1$Gyr) in
large galaxies \citep[e.g.,][Weiner et al., in
  preparation]{searle_etal:73,juneau_etal:05,willmer_etal:05}.  One implication
of this is that the total stellar mass accumulated in low-mass halos
will be much more sensitive to halo formation time compared to
high-mass halos.  Such a formation-time bias in the observed
properties of low-mass galaxies may not only help to explain the void
phenomenon, but will likely be important in attempts to construct
conditional luminosity functions that extend to low-luminosity
galaxies.  The strong clustering of early-forming low-mass halos may
also play a part in the observed trend that dim red galaxies are
substantially more clustered than their intermediate luminosity
counterparts \citep[e.g.][]{norberg_etal:02, hogg_etal:03}, although
the basic effect can be explained if the the majority of these
galaxies are satellites \citep{berlind_etal:05}.  Of course, any model
which forms galaxies using formation histories from a N-body
simulation will include the halo clustering effects implicitly, but
the extent to which this effects galaxy clustering will depend on the
efficiency of galaxy formation in low-mass halos.

There is another interesting implication of our results.  Numerous
observational signatures indicate that the central densities and
concentrations of low surface-brightness (LSB) galaxies are
significantly lower than the standard \LCDM\ paradigm predicts
\citep[e.g.,][]{bosch_etal:00,debattista_sellwood:00,keeton:01,
vandenbosch_swaters:01,alam_etal:02,zentner_bullock:02,zentner_bullock:03,
mcgaugh_etal:03,vandenbosch_etal:03,kuhlen_etal:05,simon_etal:05}.  The
dependence of clustering on halo concentration may also have
implications for the interpretation of the concentrations of LSBs.
LSBs have been shown to be notably less clustered than typical
galaxies (\citealt{mo_etal:94, rosenbaum_bomans:04}, but see also
\citealt{peebles:01}).  This might imply that they reside in a biased
population of late-forming, low-concentration halos.  If the typical
host halos for these galaxies are around $\sim 0.01 \mstar$, and LSB
galaxies are about $60\%$ less biased than typical galaxies, this
would imply that they have concentration values that are about
$1\sigma$ below the mean.  This would reduce 
the tension between the predicted and observed
concentrations of the halos hosting these galaxies.

Finally, our results also have implications for estimates of the
cluster mass function using optically-selected cluster samples.  For a
richness-selected sample of clusters, or any sample where the primary
observable that clusters are selected on is correlated with formation
time or halo concentration, mass estimates from ``self-calibration'',
using the clustering of clusters \citep{majumdar_mohr:04, lima_hu:05},
may bias results towards higher masses.  Note that because these
effects correlate with concentration, and probably halo shape and
merger history, they could affect SZ- or X-ray-selected samples as
well.  Similar biases could potentially manifest in weak lensing
measurements, and could lead to over-estimates of the mass-to-light
ratios in methods that use clustering to constrain the HOD or the
conditional luminosity function \citep[e.g.][]{bosch_etal:03}.  They
could also impact clustering-based mass estimation for other
populations that live in high $M/M_*$ halos, e.g., high-redshift bright
galaxies or quasars.  

Further work is needed to make quantitative estimates of these effects
at high masses, but this trend may be measurable in current optical
cluster samples (e.g., from the SDSS, Koester et al., in preparation).
The trend between halo occupation and formation time implies that
clusters with a given number of galaxies will be a mix of
high-concentration, early-forming, high-mass halos and
low-concentration, late-forming, low-mass halos.  If one can find an
observable measure that correlates with formation time (for example,
the difference between the luminosity of the first and second
brightest cluster galaxies, or the star formation histories of the
satellite galaxies), these correlations make specific predictions.  At
fixed richness or $\Ngal$, the early-forming sample should be more
massive and more concentrated (because concentration varies much more 
slowly with halo mass than with formation time: $c \sim m^{0.1}$, \citealt{bullock_etal:01}),
but less clustered than expected for typical halos of that mass.

Our results indicate that the Universe is somewhat more complicated
than our simplest models. However, the complication should be viewed
as an opportunity rather than an obstacle, as we can potentially learn
a great deal about details of galaxy formation in halos and their
evolutionary histories from the trends discussed in this work.  The
current and upcoming large galaxy surveys (e.g., SDSS,
\citealt{adelman_etal:06}; DEEP2, \citealt{coil_etal:04}; DES,
\citealt{abbot_etal:05}; LSST, and SNAP, \citealt{aldering_etal:04}) should be able to accurately
evaluate such effects and test the predicted trends.

\acknowledgments 

The authors thank Manoj Kaplinghat and Jeremy Tinker for useful
discussions, the referee, Andreas Berlind for several helpful comments
that improved the paper, and P. Rogers Nelson for inspiration
throughout.  The simulations were run on the Columbia machine at NASA
Ames and on the Seaborg machine at NERSC (Project PI: Joel Primack).
We thank Anatoly Klypin for running these simulations and making them
available to us.  RHW is supported by NASA through Hubble Fellowship
grant HST-HF-01168.01-A awarded by the Space Telescope Science
Institute.  ARZ is funded by the Kavli Institute for Cosmological
Physics at The University of Chicago and by the National Science
Foundation under grant NSF PHY 0114422. JSB is supported by NSF grant
AST-0507916 and by the Center for Cosmology at UC Irvine.  AVK is
supported by the NSF under grants No.  AST-0206216 and AST-0239759, by
NASA through grant NAG5-13274, and by the KICP.  This work made use of
the NASA Astrophysics Data System.


\bibliography{env}
\end{document}